\documentclass[aps,prl,twocolumn,showpacs,superscriptaddress,reprint]{revtex4-1}
\usepackage[hidelinks]{hyperref} 
\usepackage{units,float}
\usepackage[hidelinks]{hyperref} 
\usepackage{cleveref}
\usepackage{comment}
\usepackage{dsfont}
\usepackage{amsthm}
\usepackage{amsmath}
\usepackage{amssymb}
\usepackage{lipsum}
\usepackage{esint}
\usepackage{graphicx, accents}
\usepackage{mathrsfs}             
\usepackage{amssymb}
\usepackage{epstopdf}
\usepackage{textpos}
\usepackage{lipsum}
\usepackage{color}
\usepackage{blindtext}
\usepackage{braket}
\usepackage{bbm}

\begin{document}
\title{Attaining the quantum limit of super resolution in imaging an object's length via pre-detection spatial mode sorting}

\author{Zachary Dutton}
\affiliation{Physical Sciences and Systems, Raytheon BBN Technologies, Cambridge, MA 02138}
\author{Ronan Kerviche}
\affiliation{College of Optical Sciences, University of Arizona, Tucson, AZ 85719}
\author{Amit Ashok}
\affiliation{College of Optical Sciences, University of Arizona, Tucson, AZ 85719}
\author{Saikat Guha}
\affiliation{College of Optical Sciences, University of Arizona, Tucson, AZ 85719}
\affiliation{Physical Sciences and Systems, Raytheon BBN Technologies, Cambridge, MA 02138}

\begin{abstract}
Recent work considered the ultimate (quantum) limit of the precision of estimating the distance between two point objects. It was shown that the performance gap between the quantum limit and that of ideal continuum image-plane direct detection is the largest for highly sub-Rayleigh separation of the objects, and that a pre-detection mode sorting could attain the quantum limit. Here we extend this to a more general problem of estimating the length of an incoherently radiating extended (line) object. We find, as expected by the Rayleigh criterion, the Fisher information (FI) per integrated photon vanishes in the limit of small length for ideal image plane direct detection. Conversely, for a Hermite-Gaussian (HG) pre-detection mode sorter, this normalized FI does not decrease with decreasing object length, similar to the two point object case. However, unlike in the two-object problem, the FI per photon of both detection strategies gradually decreases as the object length greatly exceeds the Rayleigh limit, due to the relative inefficiency of information provided by photons emanating from near the center of the object about its length. We evaluate the quantum Fisher information per unit integrated photons and find that the HG mode sorter exactly achieves this limit at all values of the object length. Further, a simple binary mode sorter maintains the advantage of the full mode sorter at highly sub-Rayleigh length. In addition to this FI analysis, we quantify improvement in terms of the actual mean squared error of the length estimate. Finally, we consider the effect of imperfect mode sorting, and show that the performance improvement over direct detection is robust over a range of sub-Rayleigh lengths.
\end{abstract}
\maketitle

\section{Introduction}

In optical imaging, Rayleigh's criteria~\cite{Rayeligh:79} asserts that it is impossible to resolve features of the scene whose angular extent is smaller than $\lambda / D$, where $\lambda$ is the wavelength of light and $D$ is the length of the receiver pupil. This is often known as the diffraction limit. Subsequently more accurate treatments found that estimating the angular separation between two closely spaced point objects is possible even as the separation falls below the diffraction limit, by collecting a growing fraction of photons (by increasing the integration time) thereby improving the signal-to-noise ratio (SNR), as the separation decreases towards zero~\cite{Bettens:99}. However, the Fisher Information (FI), which governs the resolvability (inverse of the FI is a lower bound to an unbiased estimator of the separation), normalized by the total collected mean photon number over the integration time, does degrade as the separation falls below the diffraction limit, vanishing to zero at zero separation. Very recently, however, it was found that the aforesaid degradation of FI per unit collected photons to zero at zero angular separation is a mere artifact of image-plane (intensity) detection. Remarkably, the Quantum Fisher Information (QFI)---the highest FI achievable with {\em any} physically-permissible optical detection scheme---per unit collected photons, remains constant as the separation between the objects shrinks to zero~\cite{Tsang:16}. Calculating the QFI only requires us to specify the quantum state of the (in this case, classical thermal) light collected over the camera's integration time. But, for all single-parameter estimation problems, there must always exist an actual detection scheme whose FI exactly matches the QFI. Tsang {\em et al.} also found that for this problem, the QFI can be explicitly achieved by photon detection of the image-plane Hermite-Gaussian (HG) modes, when imaging with a Gaussian point spread function (PSF)~\cite{Tsang:16}. When a standard hard aperture pupil is employed, the QFI is attained by spatial-mode-resolved photon detection on the image-plane sinc-Bessel modes~\cite{Kerviche:17}.

Since a linear mode transformation on an optical field is in principle a reversible operation, if optical detection---conversion of the optical signal into an electrical signal, such as a photocurrent---were noiseless, a mode transformation would not alter the amount of information in the detector output. The primary intuition behind the result of~\cite{Tsang:16} and ours in this paper is that there is a minimum inevitable detection noise because of the laws of physics; and an optical-domain mode transformation can pre-dispose the information-bearing optical field in a more information favorable way to the detection noise that is yet to come. In fact, with shifted PSFs from sub-Rayleigh-separated sources in the image plane, intensity detection is one of the worst choices of optical detection to resolve the sources. 

There have been several recent works looking into practical realizations of this improved resolution.  For example, while the optimal QFI could be achieved by implementing a spatial-mode demultiplexer (SPADE), it was shown that that implementing a simpler binary SPADE detector, in which only the zeroth or the first HG mode is separately detected from its orthogonal complement (i.e., employing just two detectors), one can achieve nearly the same quantum-optimal performance in the regime of sub-Rayleigh separations. Other proposals for beating the direct imaging limit have included Super Localization by Image inVERsion interferometry (SLIVER)~\cite{Nair:16}, in which even and odd components of the image are interferometrically separated and detected, and methods to utilize the freedom in the local oscillator mode of homodyne and heterodyne detection to selectively image specific modes~\cite{Yang:16, Yang:17}. Several experimental realizations have also been recently reported~\cite{Tham:16, Tang:16}.

\begin{figure}
\centering
\includegraphics[width=\columnwidth]{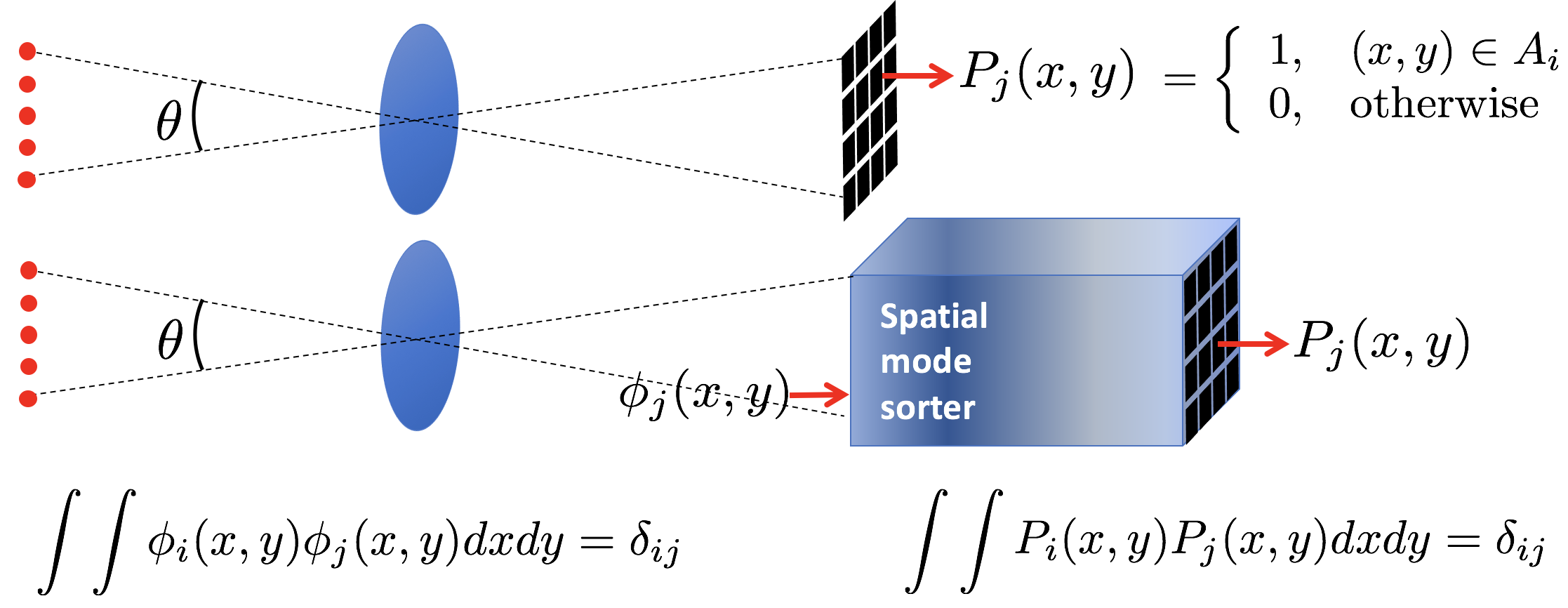}
\caption{A schematic of an incoherently emitting linear object---modeled as a collection of $M$ equal-intensity equal-separated point sources, with $M \to \infty$---onto a detector array (top) versus a spatial mode demultiplexer (SPADE) which manipulates the image-plane optical field before photon counting, i.e., detecting intensity in a different spatial mode basis (bottom).}
\label{fig:diagram}
\end{figure}

\section{Imaging an extended object}

The goal of this paper is to extend this concept of pre-detection mode sorting for attaining the fundamental limit of super-resolution imaging to the more general scenario of imaging extended objects, that can be thought of as a continuum limit of many point sources radiating incoherently over the extent of the object. The specific problem we consider here is that of estimating the (angular) length $\theta$ of an on-axis line-of-sight one dimensional (1D) object that is incoherently emitting quasi-monochromatic light uniformly along its length (see Fig.~\ref{fig:diagram}). Our technique is in principle simple to extend to more complex object shapes. We assume a Gaussian amplitude point spread function (PSF) $A(x) = (2\pi \sigma^2)^{-1/4} {\rm exp}[-x^2/4\sigma^2]$, and let $N$ be the mean photon number of the field collected over the integration time.

\begin{figure}
\centering
\fbox{\includegraphics[width=\linewidth]{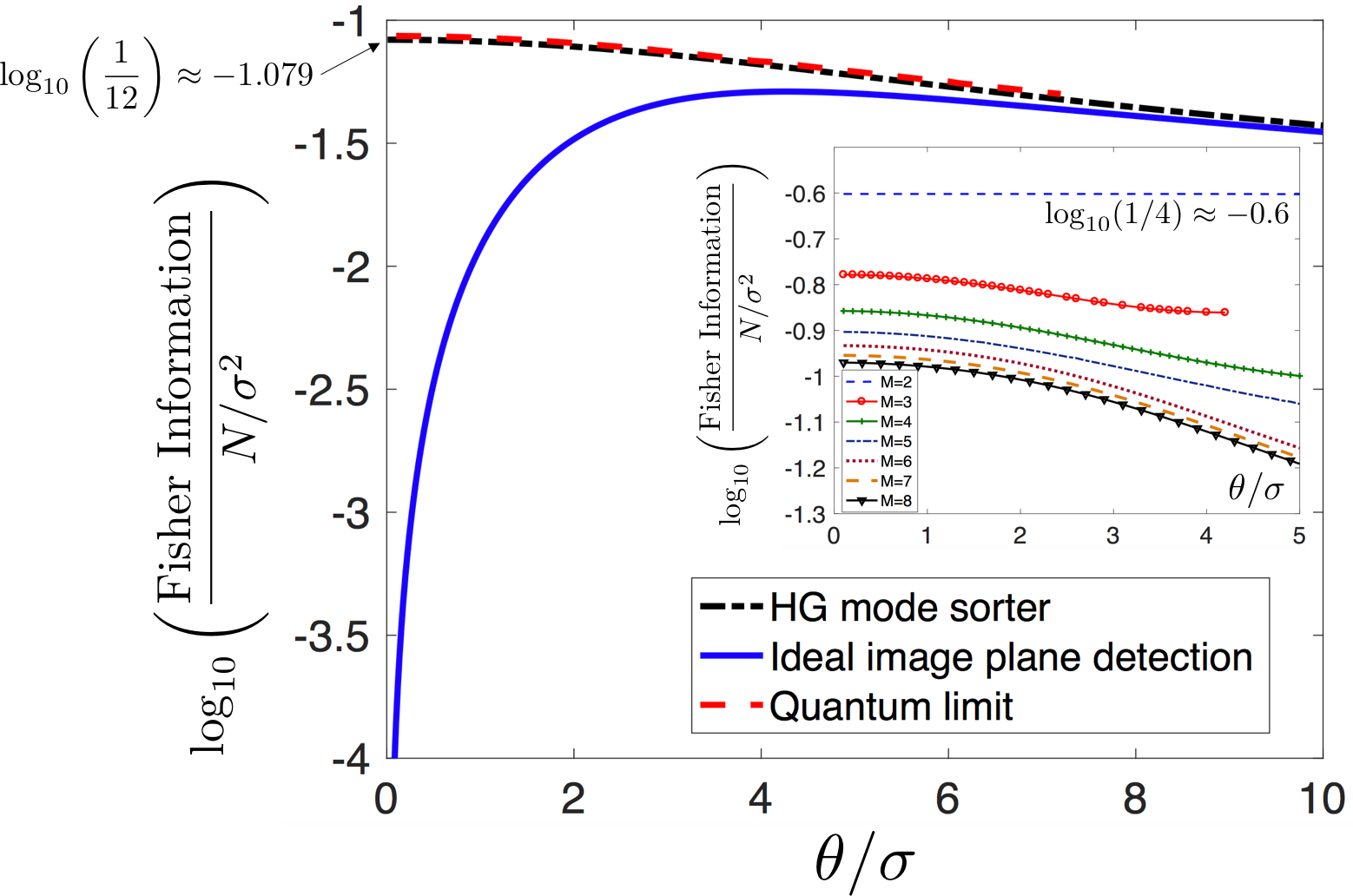}}
\caption{Fisher information (FI) per photon for estimating the length $\theta$ of an extended 1D object: Ideal continuum image plane direct detection (red dashed), infinite HG mode sorter (black dash-dotted), quantum limit (blue solid). Inset: FI per photon when the line object is approximated as $M$ equally spaced equal intensity point sources along its length.}
\label{fig:FI_compare}
\end{figure}
\subsection{Summary of main results}

In Fig.~\ref{fig:FI_compare}, we plot FI normalized by $N/\sigma^2$, as a function of $\theta/\sigma$. In analogy to the two-point source scenario~\cite{Tsang:16}, we find that: (a) FI attained by ideal image-plane detection sharply decreases towards zero as $\theta$ goes below the diffraction length ($\sigma \sim 1$) towards zero, (b) the QFI remains approximately constant as a function of $\theta$, and (c) image-plane HG-mode-resolved photon detection attains the quantum optimal performance. However, unlike the two-source case, all three normalized FI plots show a steady decline as $\theta$ increases beyond $\approx 4\sigma$. This happens because as the object's length $\theta$ grows large, a growing fraction of the photons collected by the camera comes from the inner regions of the object which carry less information about $\theta$ compared to the photons emitted from the two end points of the object; thereby diminishing the information per collected photon as $\theta$ grows large.

To understand the performance of more practical implementations we analyze the performance of binary SPADE receivers. In analogy the results for the two-point-source problem~\cite{Kerviche:17}, we find that a binary SPADE---based either on separating the HG0 mode or the HG1 mode from the respective orthogonal complement---attains the QFI as $\theta \to 0$. So, most of the advantage of the infinite HG mode sorter is maintained using just a binary mode sorter for $\theta$ much below the diffraction length, and this advantage is reasonably robust to photon leakage across desired modes in an imperfect mode-sorter implementation (see Fig.~\ref{fig:relative_FI}). We also analyze mean square error (MSE) estimates via numerical simulations, and show that the performance advantage of a mode sorter based receiver over image-plane direct detection increases as the total optical energy collected increases. This is in analogy to a recent result studying homodyne-mode-selective receiver implementations for estimating two-point-source separation~\cite{Yang:17}.

\subsection{Derivations of receiver Fisher Informations and the QFI}

For simplicity, we will assume a Gaussian PSF throughout this paper. However we expect our results to readily generalize to other PSFs~\cite{Kerviche:17}. We will also assume that the position (e.g., of the centroid) of the line object is apriori known.

Let us first consider {\em ideal} image-plane direct detection to serve as the baseline. In other words, we consider a unity fill factor focal plane array with infinitesimally small pixels of unity detection efficiency filling the entire (infinite) extent of the camera's image plane. We can express the FI as~\cite{Tsang:16}:

\begin{equation}
{\cal I}^{\text{direct}} = \frac{N}{\sigma^2} \int_{-\infty}^{\infty} \frac{1}{\Lambda(x; \theta)}\left(\frac{\partial \Lambda(x;\theta)}{\partial \theta}\right)^2 dx,
\label{eq:Idirect_integral}
\end{equation} 
where $\Lambda(x;\theta)$, the normalized ($\sigma = 1$) image-plane probability density of the photon position $x$ is given by:
\begin{eqnarray}
\Lambda(x;\theta) &=& \frac{1}{\sqrt{2\pi}}\int_{-1}^1 e^{-\left(x - \frac{\theta y}{2}\right)^2/2} dy \nonumber \\
&=& \frac{1}{\theta} \left[{\rm erf}\left(\frac{\theta - 2x}{2\sqrt{2}}\right) + {\rm erf}\left(\frac{\theta + 2x}{2\sqrt{2}}\right)\right],
\end{eqnarray}
and,
\begin{eqnarray}
\frac{\partial \Lambda(x;\theta)}{\partial \theta} = \frac{1}{\theta\sqrt{2\pi}}\left[e^{-\frac{(\theta - 2x)^2}{8}} + e^{-\frac{(\theta + 2x)^2}{8}}\right] - \frac{\Lambda(x;\theta)}{\theta}.
\end{eqnarray}

We evaluated the integral in~\eqref{eq:Idirect_integral} using Mathematica to compute ${\cal I}^{\text{direct}}$. In Fig.~\ref{fig:FI_compare}, the solid blue curve is a plot of ${\cal I}^{\text{direct}} / (N/\sigma^2)$ vs. $\theta/\sigma$. We see that the FI per photon degrades as $\theta$ becomes smaller than the Rayleigh length ($\theta / \sigma \approx 1$) and vanishes as $\theta \to 0$. In the two point source case, the FI per photon at high $\theta$ asymptotically approaches a constant~\cite{Tsang:16}. Here, by contrast, the light emitting from the center of the object communicates less information per photon compared to the light emitted from the edges, resulting in the FI per photon to diminish as $\theta$ increases beyond a certain threshold ($\sigma \approx 4.26$).


We now consider the performance of the HG mode sorter. It is straightforward to evaluate the fraction of the $N$ photons collected during the integral time that appears in the $q$-th image-plane HG mode as:

\begin{equation}
P(q;\theta) = \frac{2\gamma\left(q+\frac{1}{2},z\right)}{\theta \, q!},
\end{equation}
where $z = \theta^2/16$ and $\gamma(a,z) = \int_0^z t^{a-1}e^{-t} dt$ is the un-normalized incomplete gamma function. We define
\begin{eqnarray}
Q(q;\theta) &=& \frac{\partial}{\partial \theta} \ln P(q;\theta) = \frac{1}{P(q;\theta)} \frac{\partial}{\partial \theta} P(q;\theta) \nonumber \\
&=& \frac{1}{P(q;\theta)}\frac{e^{-z}z^{q-\frac{1}{2}}}{4q!} - \frac{1}{\theta}.
\end{eqnarray}
A receiver which then detects photons (with shot noise limited precision) on each of those modes would yield a FI given by: 
\vspace{-2pt}
\begin{equation}
{\cal I}^{\text{HG}} = \frac{N}{\sigma^2}\sum_{q=0}^\infty P(q;\theta) Q(q;\theta)^2.
\end{equation} 
\vspace{-1pt}

The black dash-dotted curve in Fig.~\ref{fig:FI_compare} plots ${\cal I}^{\text{HG}}/(N/\sigma^2)$. We see that for small (sub-Rayleigh) $\theta$, the HG basis measurement performs far better than ideal direct imaging and the FI per photon does not degrade as $\theta$ becomes small. At high $\theta$ the FI per photon is seen to asymptotically converge to that of ideal image plane detection. We can prove that $\lim_{\theta \to 0}{\cal I}^{\rm{HG}} = N/12\sigma^2$.

Finally we wish to calculate the quantum Fisher information (QFI), the optimal performance attainable by any receiver. We follow a procedure similar to~\cite{Tsang:16}, but model the extended object as a collection of $M$ equally spaced point emitters spanning the total angular length $\theta$, each radiating incoherently but within a narrow band of $W$ Hz around a center wavelength $\lambda$, and then take the limit of $M \to \infty$. Over the integration time $T$, there are roughly $K \approx WT$ orthogonal temporal modes. At optical frequencies, the mean photon number per mode ${\bar n} \ll 1$, and hence one can express the density operator of the entire collected optical field as $\omega = \rho^{\otimes K}$, where $\rho$, the state of each mode can be written as: $\rho = (1-{\bar n})\rho_0 + {\bar n}\rho_1 + O({\bar n}^2)$, where $\rho_0 = |{\rm{vac}}\rangle \langle {\rm{vac}}|$ is the multi-spatio-temporal-mode vacuum state, and $\rho_1$ is a single-photon mixed state, $\rho_1 = \int_{-\theta/2}^{\theta/2}|\psi_x\rangle \langle \psi_x | dx$, where the pure state $|\psi_x\rangle$ is that of a single photon in a shifted image-plane amplitude-PSF spatial mode centered at $x$. The total photon number $N = {\bar n}K$. In the above model, ${\bar n}$ can also be interpreted as the probability that a temporal mode of the collected light (over all spatial modes) has one photon. The QFI, ${\cal I}^{\text{Q}}(\omega) = K {\cal I}^{\text{Q}}(\rho)$, and we use the approximation ${\cal I}^{\text{Q}}(\rho) \approx {\bar n}{\cal I}^{\text{Q}}(\rho_1)$~\cite{Tsang:16}, where ${\cal I}^{\text{Q}}(\rho_1) = {\rm Tr}(\rho_1 {\mathcal L}^2)$. The symmetric logarithmic derivative (SLD) $\mathcal{L}$ is given by the implicit relation $\partial \rho_1 / \partial \theta = \nicefrac{1}{2}\left[\rho_1{\mathcal{L}} + {\mathcal{L}}\rho_1\right]$, and can be expressed as ${\mathcal L} = \sum_{j,k; D_j + D_k \ne 0}\frac{2}{D_j + D_k}\langle e_j | \frac{\partial \rho}{\partial \theta} | e_k \rangle |e_j\rangle\langle e_k |$, where $D_j$ and $|e_j\rangle$ are the eigenvalues and eigenvectors of $\rho$, i.e, $\rho = \sum_j D_j |e_j\rangle \langle e_j|$. Next we observe that the single-photon state in the shifted amplitude PSF mode centered at $x$, $|\psi_x\rangle = \sum_{k=0}^{\infty} e^{-|\alpha|^2/2}\frac{\alpha^k}{\sqrt{k!}}|\phi_k\rangle$, where $|\phi_k\rangle$ is the state of a single photon in the $k$-th image-plane HG mode, and $\alpha = x/2\sigma$. For $M > 2$ point sources, an analytical calculation of the QFI becomes involved. We calculate the QFI numerically by first expressing $\rho_1$ in the $\left\{|\phi_k\rangle\right\}$ basis, and then calculate the eigen-basis of $\rho_1$ and the eigenvectors which span $\partial \rho_1/\partial \theta$, then $\mathcal L$, and finally the QFI as, ${\cal I}^{\text{Q}}(\omega) = K{\bar n}{\cal I}^{\text{Q}}(\rho_1)$.

As expected, the $M=2$ case reproduces the constant QFI, ${\cal I}^Q/(N/\sigma^2) = 1/4$, as obtained in Ref.~\cite{Tsang:16}. The QFI for $M = 2, 3, \ldots, 8$ sources are plotted in the inset of Fig.~\ref{fig:FI_compare}. Increasing $M$ introduces a slow degradation with higher $\theta$ as well as a suppression of the normalized QFI across all $\theta$. As we increased $M$ we eventually converged to a constant curve. We found that $M=56$ was sufficient to obtain a reliable plot that was independent of $M$, which we interpret as the QFI for the continuum object. Although we do not have an analytical proof that ${\cal I}^Q = {\cal I}^{\rm{HG}}$, $\forall \theta$, our numerical results shown in Fig.~\ref{fig:FI_compare} suggests that is true, confirming that HG-mode-resolved photon detection is the optimal receiver for this problem. The reason the plot of ${\cal I}^Q$ is slightly above that of ${\cal I}^{\rm{HG}}$ is that the $M$ value we use to evaluate (an approximation of) ${\cal I}^Q$ is finite, and exact convergence would happen in the continuum limit, $M \to \infty$.

\section{Discussion}
\begin{figure}
\centering
\fbox{\includegraphics[width=0.8\columnwidth]{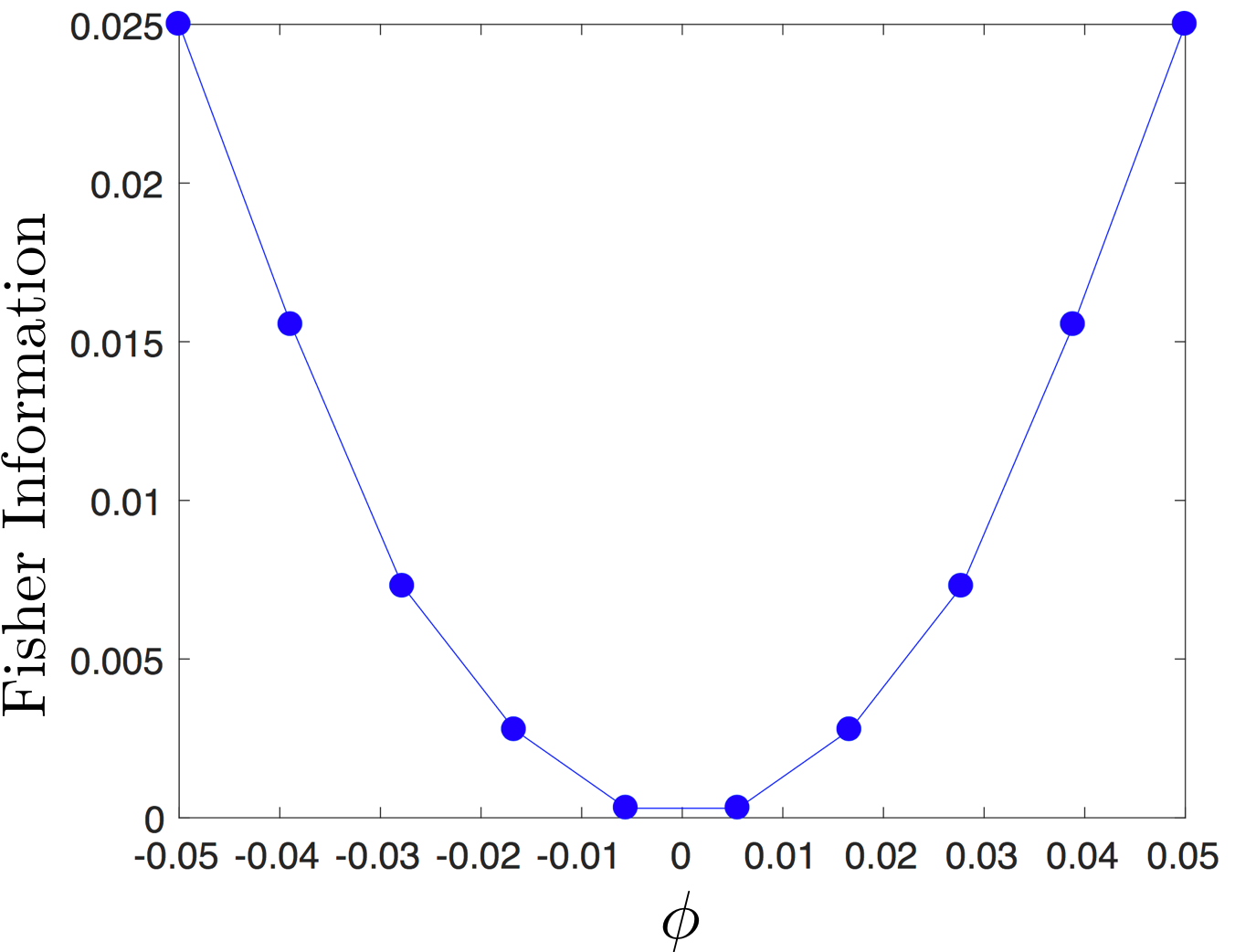}}
\caption{Fisher information per photon for estimating the angular length ($\theta$) of the object when all the photon energy is assumed to be emitted from a single point at a fixed angular location $\phi \in [-\theta/2, \theta/2]$ that is assumed to be apriori known.}
\label{fig:relative_FI}
\end{figure}
\subsection{Relative information contribution of photons emitted from different parts of the object}
In order to quantify the contribution to the FI from light emanating from the angular location $\phi \in [-\theta/2, \theta/2]$ of the object, we do the following thought experiment. Instead of uniformly distributing emitters along the entire object, we concentrate all $M$ emitters at one given (point) location $\phi$. We assume $\phi$ is known. Calculating the QFI for estimating $\theta$ as we vary $\phi$ produces the plot in Fig.~\ref{fig:relative_FI}. Here we can see the relative information inefficiency of light coming from near the center of the object ($\phi = 0$). It is the sum of this information distribution across the length of the object that results in the slow degradation of the total FI per photon for large $\theta$ as seen in all the plots in Fig.~\ref{fig:FI_compare}.

\begin{figure}[h]
\centering
\fbox{\includegraphics[width=\columnwidth]{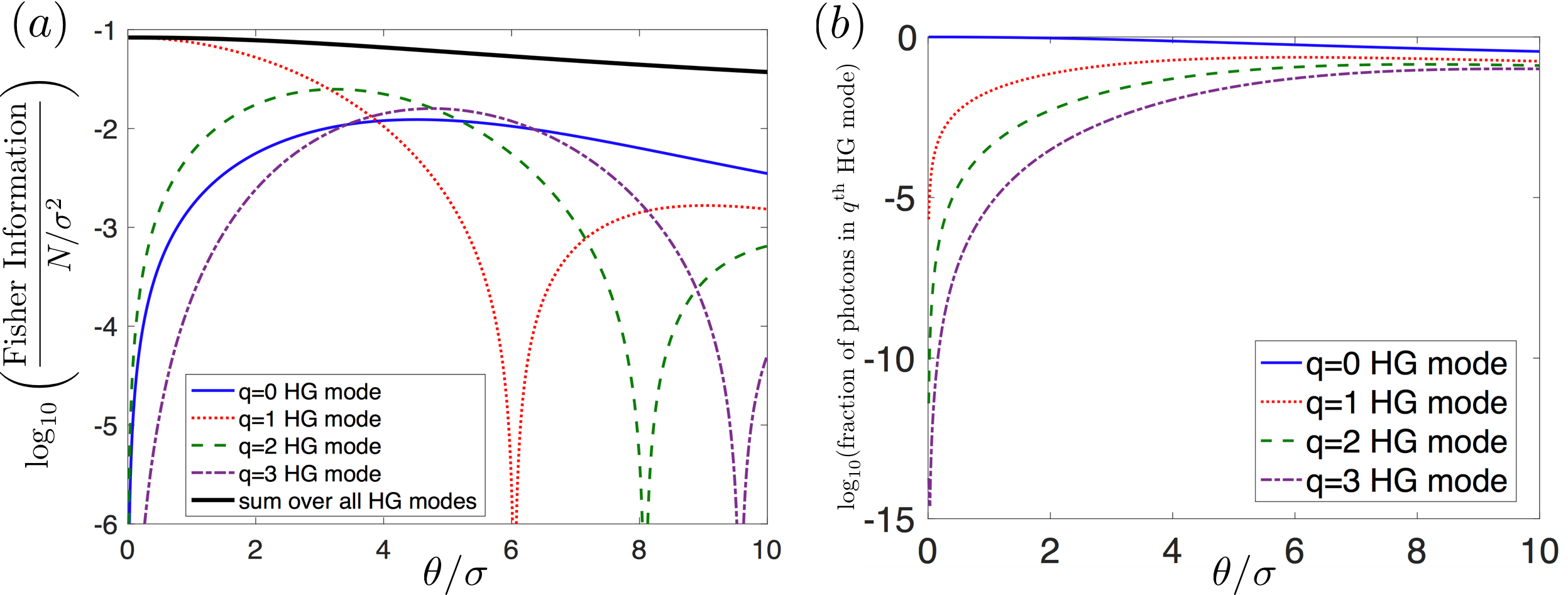}}
\caption{(a) The individual FI contributions from the $q=0,1, 2$ and $3$ modes, and the total FI from all HG modes (which equals the QFI); (b) fraction of image-plane photons in the $q^{\rm th}$ HG mode. Most of the information is contained in the $q=1$ (first antisymmetric) mode, but most of the photons are contained in the $q=0$ (PSF) mode, for small $\theta/\sigma$.}
\label{fig:FI_modes}
\end{figure}
\subsection{Modal distribution of information}

\begin{figure}[h]
\centering
\fbox{\includegraphics[width=0.9\linewidth]{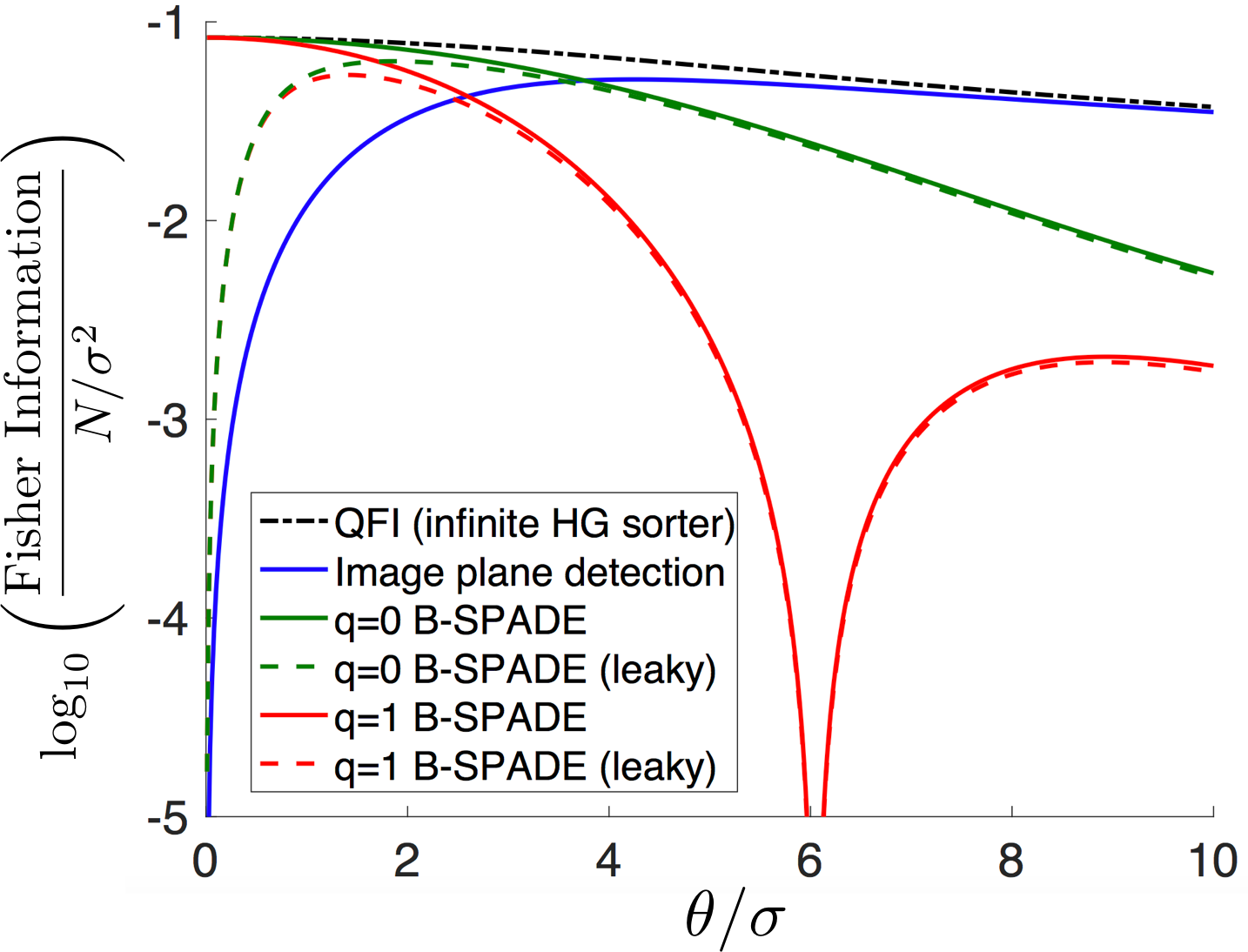}}
\caption{FI attained by a binary SPADE receiver, based on either separating the $q=0$ (green) or the $q=1$ (red) HG mode. The FI attained by the infinite HG mode sorter (which equals the QFI) and that of continuum direct detection are reproduced.  Finally, the green and red dashed curves show the performance of the $q=0$ and $q=1$ binary SPADE detectors but with leakage parameter, $\epsilon=0.01$. Photon leakage is seen to degrade the performance as $\theta$ approaches zero, but still outperforms ideal direct detection for $\epsilon=0.01$.}
\label{fig:leakysorter}
\end{figure}

To better understand the modal information composition, we plot in Fig.~\ref{fig:FI_modes} the Fisher information contributed by each of the first few individual terms in the expression ${\cal I}^{\text{HG}}/(N/\sigma^2)$. Just as in the two point source case~\cite{Kerviche:17}, the HG 1 mode carries almost all the information at low $\theta$. FI plots for each mode has a node at which the information content vanishes. A binary SPADE receiver separates all the collected light into one spatial mode and its orthogonal complement. In Fig.~\ref{fig:leakysorter} we plot the performance of this receiver where the separated mode is either the $q= 0$ mode (green plots) or the $q=1$ modes (red plots). The two perform similarly and approach the QFI at low $\theta$, but differ significantly once we approach the aforesaid nodes in the individual information plots in Fig.~\ref{fig:FI_modes} associated with $q=1$.  

\begin{figure}[h]
\centering
\fbox{\includegraphics[width=0.8\columnwidth]{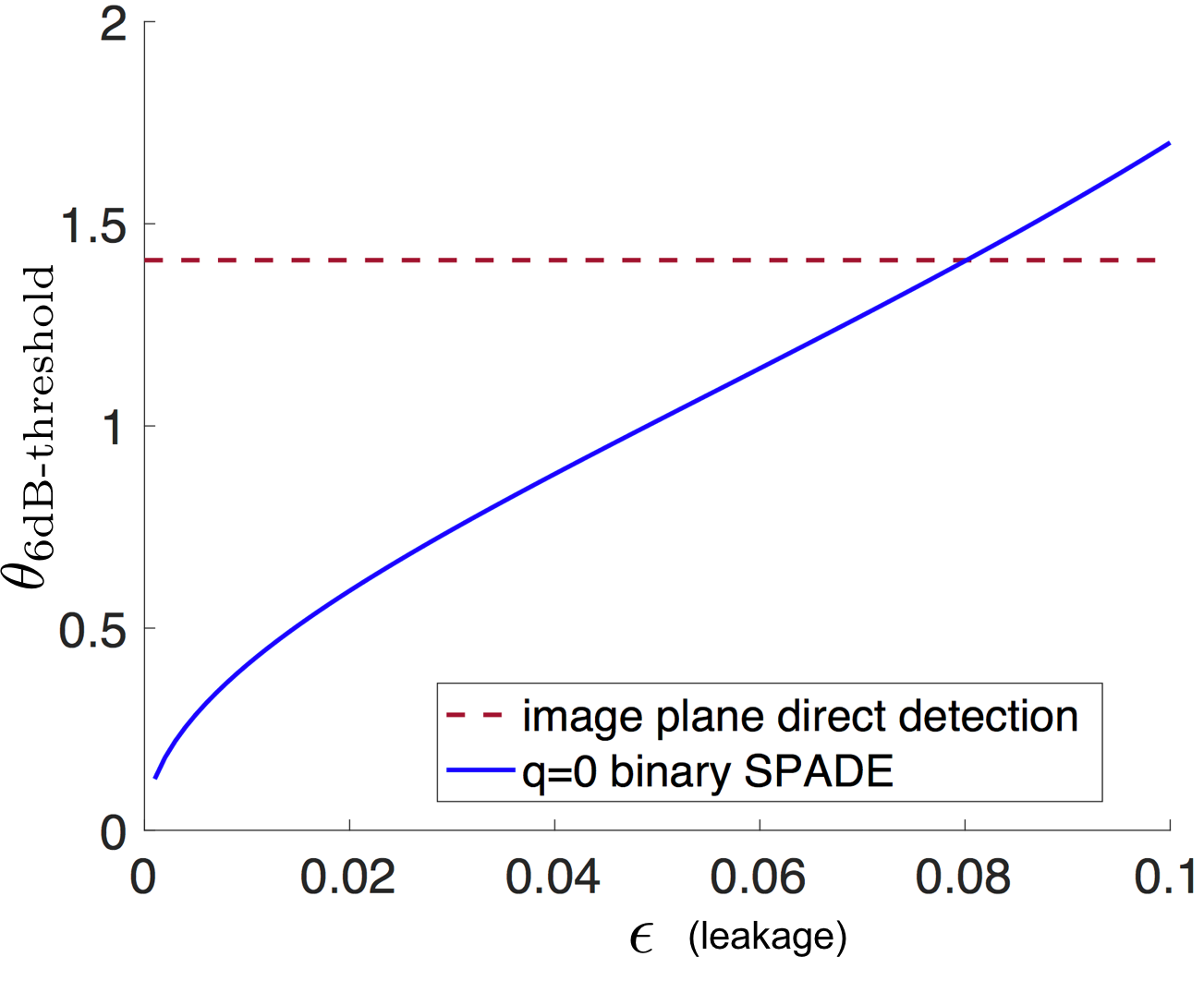}}
\caption{Plots of the $\theta$ at which the FI reaches the threshold of 6 dB below the ideal QFI at $\theta=0$.  The burgundy curve show the direct detection case ($\theta=1.4\sigma$) while the $q=0$ blue curve shows the degradation of this 6 dB point for the binary SPADE as the leakage $\epsilon$ is increased.}
\label{fig:threshold}
\end{figure}
\subsection{Effect of imperfections in mode sorter}

Next, we evaluate the sensitivity of these binary SPADE receivers to leakage in the mode-sorter implementation, and define $\epsilon$ to be the fractional power in the intended mode that leaks into the output corresponding to the orthogonal complement (and vice versa). These are plotted as dashed curves for $\epsilon=0.01$ in Fig.~\ref{fig:leakysorter}. The performance of both receivers degrade near $theta=0$ but are insensitive to leakage away from this point.  Finite leakage essentially returns the binary SPADE to a condition where the FI once again vanishes at $\theta \to 0$ but the region where it is suppressed is a function of $\epsilon$. To quantify this over a range, we plot in Fig.~\ref{fig:threshold} (blue curve) the $\theta$ where the performance of the $q=0$ binary SPADE dips to $6$ dB below the ideal HG mode sorter's $\theta=0$ performance. One sees this increases roughly linearly with $\epsilon$ and but maintains it's advantage over direct detection until $\epsilon \lesssim 0.08$.

\subsection{Mean squared error performance}
\begin{figure}[h]
\centering
\fbox{\includegraphics[width=0.9\columnwidth]{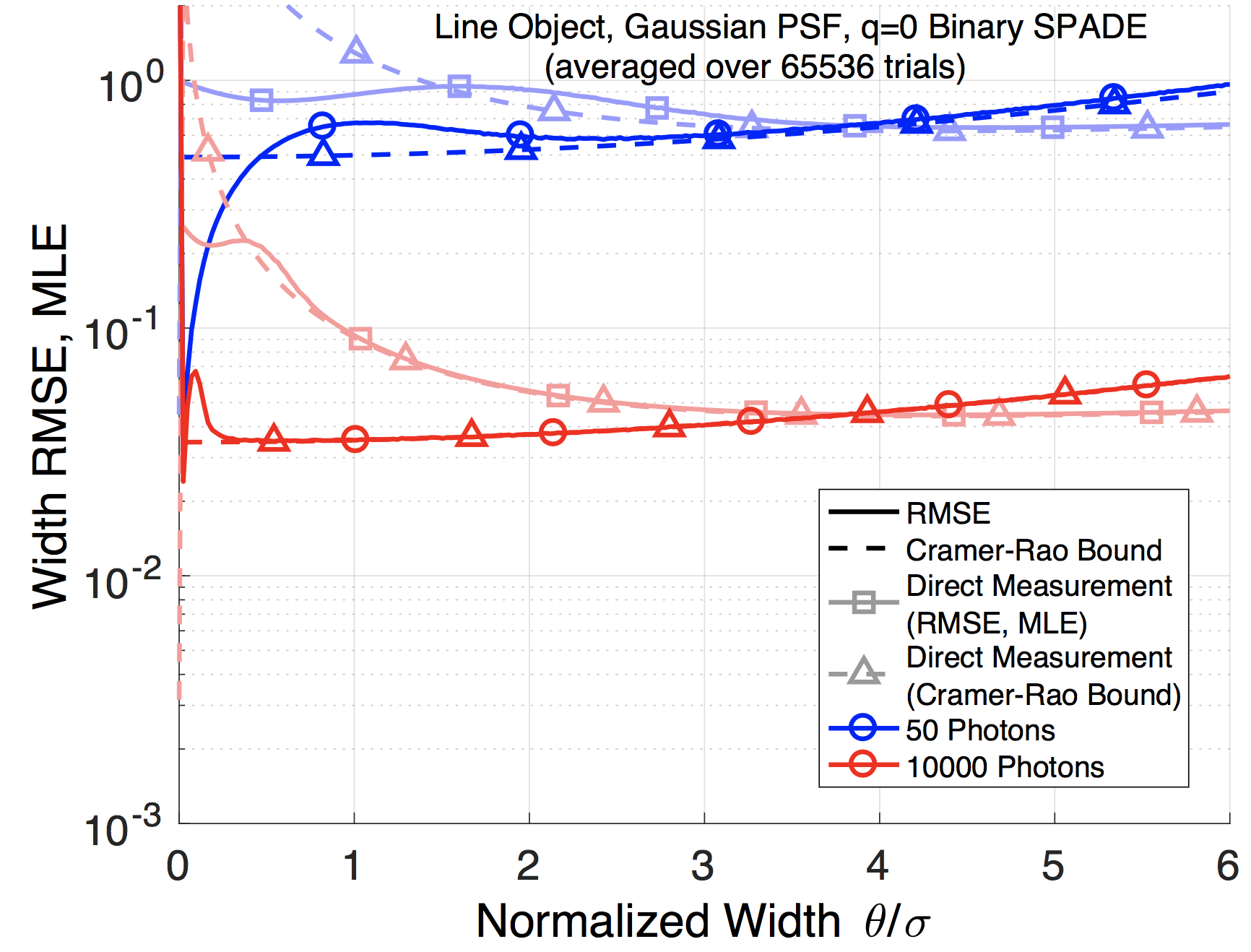}}
\fbox{\includegraphics[width=0.9\columnwidth]{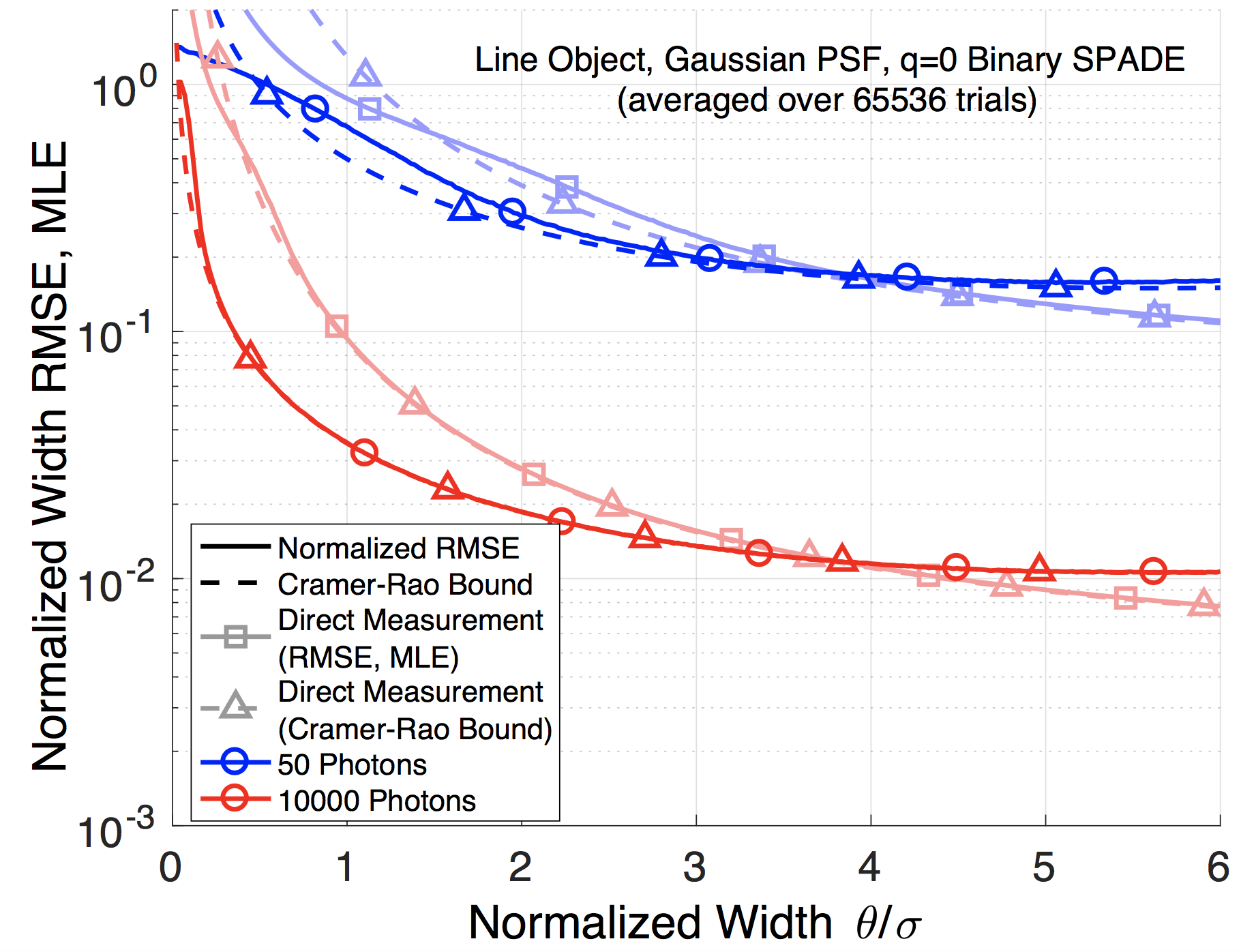}}
\caption{Comparison of RMSE (solid plots) with CR bounds (dashed plots); RMSE (top) and normalized RMSE (bottom), where RMSE in estimating $\theta$ is expressed as a fraction of the true value of $\theta$ itself. Lighter shade plots correspond to image-plane direct measurement, and darker shade plots to $q=0$ binary SPADE. Blue plots: $N=50$, red plots: $N=10,000$.}
\label{fig:RMSE}
\end{figure}

So far, we have quantified receiver performance in terms of the FI, the inverse of which---the Cramer-Rao (CR) bound---gives a lower bound on the mean squared error (MSE), i.e., the variance ${\rm Var}({\hat \theta}) = {\rm E}\left[({\hat \theta} - \theta)^2 | \theta\right]$, where $\hat \theta$ is an unbiased estimator of $\theta$ given the detector output. In this section, we report comparisons of the CR bound to the numerically-evaluated RMSE, an indicator of the practical performance attained by these receivers. The results are plotted for two different total photon numbers $N$ in Fig.~\ref{fig:RMSE}(a); $N=50$ (blue plots) and $N=10,000$ (red plots). CR bounds are plotted with dashed lines while the RMSE is plotted with solid lines. The lighter shades corresponding to image-plane direct detection and the darker ones corresponding to $q=0$ binary SPADE. One sees the nearly constant CR bound (corresponding to the nearly constant FI) for the $q=0$ binary SPADE. The RMSE agrees with the CR bound closely, showing the bound is quite tight. Direct measurement is seen to perform much worse at sub-Rayleigh lengths, as expected. Note that the relatively smaller advantage over direct detection corresponds to the degradation due to the binary SPADE (as opposed to using the full HG mode sorter), as is also seen in the FI curves in Figs.~\ref{fig:FI_modes},~\ref{fig:leakysorter}. For the blue ($N=50$ photon) curves, it is evident that the CR bound is less tight and, interestingly, the advantage of the binary SPADE is also less dramatic. This is consistent with a similar trend seen in~\cite{Yang:17} for a homodyne based mode-sorting detector for the two-source problem. Finally, in Fig.~\ref{fig:RMSE}(b) we show the same results as in (a) but now normalizing the RMSE to the actual object length $\theta$. Plotted in this way it becomes clearer that the relative precision of the length estimate is, quite intuitively, always degrading as $\theta$ shrinks to zero.  However the relative performance of the various cases is of course retained.

\subsection{Generalization to arbitrary PSF}

In analogy with the results in~\cite{Kerviche:17}, we conjecture the following holds for a general amplitude PSF $A(x/\sigma)/\sqrt{\sigma}$, with $\int_{-\infty}^{\infty} |A(x)|^2 dx = 1$. Let us consider the autocorrelation function of $A(x)$,
\begin{equation}
\Gamma_A(x^\prime) = \int_{-\infty}^{\infty}A^*(x)A(x+x^\prime)dx.
\end{equation}
Assuming $\Gamma_A(x)$ admits a Taylor series expansion near $\theta = 0$, as $\Gamma_A(x) = 1 + i\beta x - \frac{\alpha}{2}x^2 + \mathcal{O}(x^3)$, $\alpha>=0$, the QFI, as $\theta \to 0$, would converge to:
\begin{equation}
\lim_{\theta \to 0}{\cal I}^Q(\theta) = \frac{N}{3\sigma^2}(\alpha - \beta^2). 
\end{equation}
For a real-valued $A(x)$, $\beta = 0$. For a Gaussian PSF (the case considered in this paper), it is simple to verify that $\alpha = 1/4, \beta = 0$, whereas for a hard rectangular aperture, $\alpha = \pi^2/3, \beta = 0$. We have shown that, for any amplitude PSF $A(x)$, the Fisher Information (normalized to $N/\sigma^2)$ for object length estimation is exactly three times lower than that for the estimation of two point-source separation, as $\theta/\sigma \to 0$ (see Appendix~\ref{app:first}). 

Further, the evaluation of the FI per photon for the width estimation problem can be generalized to objects having any arbitrary irradiance profile by calculating its second order moment. This holds for any amplitude PSF $A(x)$ as shown in Appendix~\ref{app:second} with the example of a `triangular' intensity profile.


Finally, we believe all the qualitative conclusions drawn in this paper will hold for most well-behaved aperture functions, with the optimal modes (that attain the QFI) being the orthonormal mode set constructed using the $q^{\rm th}$ derivatives of $A(x)$, as explained in~\cite{Kerviche:17}. A formal proof of this, and extensions of the idea to more complex scenes is left open for future work.

\section{Conclusion}
In conclusion, we have expanded on recent results examining how non-standard imaging techniques based on pre-detection spatial mode sorting and related techniques can out-perform standard focal-plane intensity-based imaging, for estimation problems involving resolving object features smaller than the diffraction limit. We calculated the quantum limit of estimating the length of an incoherently radiating 1D extended object and found that a receiver based on sorting HG modes attains the quantum limit. We found that a binary SPADE technique, based on separating one mode from its orthogonal complement, still outperforms infinite-resolution infinite-pixelated direct imaging, and maintain near quantum optimality when the object length is in the highly sub-Rayleigh regime. We have investigated how leakage in the mode-sorter implementation degrades the performance of binary SPADE and found that its advantage over direct imaging is somewhat robust to that leakage. We have also investigated with numerical evaluation how the Fisher Information based advantage translates to an advantage in the actual RMSE attained by the mode-sorting receivers. 

Two important directions of future work are, to investigate: (1) optimal joint estimation of multiple scene parameters in more complex scenes, and (2) single-parameter estimation while treating unknown (nuisance) parameters. It is important to do a fair comparison with standard image-plane detection where both schemes are given the same integration time and the same initial priors over all scene parameters of interest and nuisance parameters. This is particularly important since all the scene information are inferred in electronic-domain post-processing in image plane detection, due to which, as an example, it does not need to need to estimate the center of an object to estimate its length, whereas our scheme would require the estimate of the center to `point' the mode sorter accurately.

\section*{Acknowledgements}
This work was supported in part by the DARPA REVEAL program under contract number HR0011-16-C-0026, and in part by a DARPA EXTREME program seedling under contract number HR0011-17-1-0007. The content or the information presented in this paper does not necessarily reflect the position or policy of either DARPA or the US Government, and no such official endorsement should be inferred.


\appendix

\section{Ratio of Fisher Information per photon: line object vs. two point source}\label{app:first}
In this Appendix, we will prove that for any amplitude PSF, the Fisher Information (normalized to $N/\sigma^2)$ for object length estimation is exactly three times lower than that for the estimation of two point-source separation, as $\theta/\sigma \to 0$. Let us consider an imaging system casting an arbitrary (possibly complex valued) amplitude PSF $A(x/\sigma)/\sqrt{\sigma}$ that is infinitely differentiable, where $\sigma$ is a positive scaling factor and such that the profile is normalized in energy, i.e., $\int_{-\infty}^{\infty} |A(x)|^2\ dx=1$, then we have shown in \cite{Kerviche:17} that an \textit{ad-hoc} mode-sorter can be devised upon the successive spatial derivatives of $A(x)$.
	
	In this particular case, we have expressed the output functions or mode projections $(m_{A,q}(\theta))$ regarding the measurement of the separation between two incoherent and indinstiguishable point sources located respectively at $-\theta/2\sigma$ and $+\theta/2\sigma$ from the autocorrelation function $\Gamma_A$ of the PSF : $\Gamma_A(x')=\int_{-\infty}^{+\infty} \overline{A(x)} A(x+x')\ dx$. Furthermore if this autocorrelation function $\Gamma_A$ admits a Taylor expansion near $\theta=0$ in the form of :
	\begin{align}
		\Gamma_A(x) \underset{\theta\rightarrow 0}{=} 1 + i\beta x - \frac{\alpha}{2} x^2 + o(x^2),
	\end{align}
	where $\alpha\ge0$, and $\beta\in\mathbb{R}$ then the Fisher Information score for narrow separation, i.e., small values of $\theta$, obtained by the second mode measurement $m_{A,1}$ tends toward $\mathcal{I}_1(\theta) \rightarrow N(\alpha-\beta^2)/\sigma^2$. In addition, we have the following expansion of the measurement function for the same second mode, which corresponds to the first spatial derivative of the PSF:
	\begin{align}
		m_{A,1}(\theta) &\underset{\theta\rightarrow0}{=} (\alpha-\beta^2)\frac{\theta^2}{\sigma^2} + o\left(\frac{\theta^2}{\sigma^2}\right)
	\end{align}
If we now consider an incoherent one-dimensional line object spanning from $-\Delta/2\sigma$ to $+\Delta/2\sigma$, the equivalent measurement functions for the width $M_q(\Delta)$ can be directly derived from the previous $(m_{A,q}(\theta))$ functions :
	\begin{align}
		M_{A,q}(\Delta) = \frac{2}{\Delta}\int_{0}^{\Delta/2} m_{A,q}(\theta)\ d\theta,
	\end{align}
	where the pre-factor $2/\Delta$ ensures that the energy collected from the object in the image plane of the device is constant and equal to one. By employing the previous expansion again, we can write for the measurement funtion of the second mode ($q=1$) :
	\begin{align}
		M_{A,1}(\Delta) \underset{\Delta\rightarrow0}{=} \frac{\alpha-\beta^2}{12}\frac{\Delta^2}{\sigma^2} + o\left(\frac{\Delta^2}{\sigma^2}\right)
	\end{align}
The Fisher Information $\mathcal{I}_q(\Delta)$ for the mode $q$ is given by :
	\begin{align}
		\mathcal{I}_q(\Delta) = N \frac{M_{A,q}^{(1)}(\Delta)^2}{M_{A,q}(\Delta)},
	\end{align}
	with $M_{A,q}^{(1)}$ being the first derivative of the measurement function for mode $q$ with respect to $\Delta$ and $N$ the total number of photons collected. Finally, all the Fisher Information is collected solely by the second mode when $\Delta\rightarrow0$ (as discussed in~\cite{Kerviche:17}) and approaches:
	\begin{align}
		\mathcal{I}_1(\Delta) \underset{\Delta\rightarrow0}{\rightarrow} \frac{N}{3\sigma^2}(\alpha-\beta^2),
	\end{align}
	which is exactly three times lower than that for the two point-sources separation estimation problem regardless of the PSF spatial profile $A(x)$.

\section{Fisher Information per photon: arbitrary incoherent object}\label{app:second}
In this Appendix, we will further generalize the expression of the mode-measurements for any arbitrary PSF $A(x)$ subject to the same constraints as before, and knowing the mode-measurements functions $(m_{A,q})$ for the estimation of the separation of two point-sources. Here, we consider a generic extended and incoherent object described by a unidimensional profile $V(x)$ of overall width $1$ : $|x| > 1/2\Rightarrow V(x)=0$ and such that : $\int_{-1/2}^{1/2} V(x)\ dx=1$. One can note that the intensity profile $V(x)$ can be expanded into the unique sum of an odd and an even profile noted respectively $O(x)$ and $E(x)$, and the even term is always positive.

	The generalized mode-measurement function can then be expressed as :
	\begin{align}
		M_{A,q}(\Delta) &= \left.\int_{-\Delta/2}^{\Delta/2} V\left(\frac{\theta}{\Delta}\right) m_{A,q}(\left| \theta \right|)\ d\theta \right/ \int_{-\Delta/2}^{+\Delta/2} V\left(\frac{x}{\Delta}\right)\ dx \\ 
		&= \frac{2}{\Delta} \int_{0}^{\Delta/2} E\left(\frac{\theta}{\Delta}\right) m_{A,q}(\theta)\ d\theta
	\end{align}

	The expansion of the second mode-measurement $M_{A,1}(\Delta)$ in the vicinity of $\Delta\rightarrow0$ is then relatable to the second order moment of the even portion of the intensity profile :
	\begin{align}
		M_{A,1}(\Delta) \underset{\Delta\rightarrow0}{=} 2(\alpha-\beta^2) \left(\int_0^{1/2} \theta^2 E(\theta)\ d\theta\right) \frac{\Delta^2}{\sigma^2} + o\left(\frac{\Delta^2}{\sigma^2}\right)
	\end{align}
	and the corresponding Fisher Information has the following limit as the width of the object tends to zero :
	\begin{align}
		\mathcal{I}_1(\Delta) \underset{\Delta\rightarrow0}{\rightarrow} \frac{8N}{\sigma^2}(\alpha-\beta^2) \int_0^{1/2} \theta^2 E(\theta)\ d\theta
	\end{align}

One can verify that for a uniform rectangular object we have $E(\theta)=1$, the second order moment over $[0;1/2]$ is equal to $1/24$ and we find the previous value of the Fisher Information. In the case of a triangular intensity profile, the even function is $E(\theta)=2-4\theta$ when $0\le\theta\le1/2$ and $E(\theta)=2+4\theta$ when $-1/2\le\theta<0$, the second order moment over $[0;1/2]$ is equal to $1/48$ and the corresponding Fisher Information is : $N(\alpha-\beta^2)/6\sigma^2$, i.e., two times smaller than for the uniform object and six times smaller than for the two point-sources separation.

\end{document}